\documentclass[preprint2]{aastex61}

\shorttitle{Explosive events and network jets}
\shortauthors{Chen et al.}

\begin{document}

\title{Investigating the Transition Region Explosive Events and Their Relationship to Network Jets}

\correspondingauthor{Hui Tian}
\email{huitian@pku.edu.cn}

\author{Yajie Chen}
\affil{School of Earth and Space Sciences, Peking University, Beijing 100871, China}

\author{Hui Tian}
\affiliation{School of Earth and Space Sciences, Peking University, Beijing 100871, China}

\author{Zhenghua Huang}
\affiliation{Shandong Provincial Key Laboratory of Optical Astronomy and Solar-Terrestrial Environment, Institute of
Space Sciences, Shandong University, Weihai, 264209 Shandong, China}

\author{Hardi Peter}
\affiliation{Max Planck Institute for Solar System Research, Justus-von-Liebig-Weg 3, 37077, G\"{o}ttingen, Germany}

\author{Tanmoy Samanta}
\affiliation{School of Earth and Space Sciences, Peking University, Beijing 100871, China}

\begin{abstract}

Recent imaging observations with the Interface Region Imaging Spectrograp (IRIS) have revealed prevalent intermittent jets with apparent speeds of 80--250 km~s$^{-1}$ from the network lanes in the solar transition region (TR). On the other hand, spectroscopic observations of the TR lines have revealed the frequent presence of highly non-Gaussian line profiles with enhanced emission at the line wings, often referred as explosive events (EEs). Using simultaneous imaging and spectroscopic observations from IRIS, we investigate the relationship between EEs and network jets. We first identify EEs from the Si~{\sc{iv}}~1393.755 {\AA} line profiles in our observations, then examine related features in the 1330 {\AA} slit-jaw images. Our analysis suggests that EEs with double peaks or enhancements in both wings appear to be located at either the footpoints of network jets, or transient compact brightenings. These EEs are most likely produced by magnetic reconnection. We also find that EEs with enhancements only at the blue wing are mainly located on network jets, away from the footpoints. These EEs clearly result from the superposition of the high-speed network jets on the TR background. In addition, EEs showing enhancement only at the red wing of the line are often located around the jet footpoints, possibly caused by the superposition of reconnection downflows on the background emission. Moreover, we find some network jets that are not associated with any detectable EEs. Our analysis suggests that some EEs are related to the birth or propagation of network jets, and that others are not connected to network jets.

\end{abstract}


\section{Introduction} \label{sec:intro}

Explosive events (EEs) are small-scale dynamic events that are characterized by non-Gaussian profiles of emission lines formed in the solar transition region (TR, e.g. Si~{\sc{iv}} 1394 {\AA}, O~{\sc{iv}} 1032 {\AA}, C~{\sc{iv}} 1548/1550 {\AA}). The line profiles often reveal enhanced emission at the wings, which generally correspond to Doppler velocities of 50--200 km~s$^{-1}$. These events were first discovered by the High Resolution Telescope Spectrograph \citep[HRTS,][]{Bartoe1975} rocket experiment, and were named as turbulent events \citep{Brueckner1983}. Most studies of EEs focus on the quiet sun or coronal holes. Explosive events often have a size of 2$^{\prime\prime}$--5$^{\prime\prime}$ and a typical lifetime of $\sim$60 s \citep[e.g.,][]{Dere1989, Teriaca2004, Innes1997,Peter2003}. Most EEs occur in or near network lanes \citep{Dere1994}. 

Explosive events are usually associated with magnetic flux cancellation in the photosphere \citep{Dere1991, Chae1998, Huang2014, Gupta2015, Samanta2015}, and the wing enhancement indicates bi-direction reconnection jets with speeds of 50--200 km~s$^{-1}$ \citep[e.g.,][]{Innes1997,Madjarska2004}. EEs have been reproduced as a result of magnetic reconnection in numerical simulations as well \citep[e.g.,][]{Innes1999, Roussev2001a, Roussev2001b, Roussev2001c, Roussev2002,Innes2015}. Observations often reveal repeatedly occurring EEs \citep{Chae1998, Ning2004, Doyle2006}, which might be caused by magnetic reconnection in the upper chromosphere modulated by p-mode oscillations \citep{Chen2006}.

Though EEs are often believed to be manifestations of small-scale reconnection events, non-Gaussian line profiles could be generated due to other effects, such as spinning, unwinding, or twisting motions \citep{Curdt2011,Curdt2012}. \citet{DePontieu2014b} have demonstrated that twisting motions of small-scale TR loops and jets can result in EE-type line profiles. \citet{Huang2014} found an EE associated with plasma ejection followed by retraction in the chromosphere.

Recently, \citet{Tian2014} reported prevalent intermittent jets from network lanes based on imaging observations by the Interface Region Imaging Spectrograph \citep[IRIS, ][]{DePontieu2014}. These network jets can be identified from the slit-jaw images (SJI) taken with the 1330/1400 {\AA} filters, which sample plasma with typical TR temperatures. The apparent velocities of these small-scale jets are usually in the range of 80--250 km~s$^{-1}$. The lifetimes of these jets are often less than one minute. The network jets occur in both coronal holes and quiet sun regions, and those in coronal holes appear to be longer and faster \citep{Narang2016}. 

The generation mechanisms of these jets are still under investigation. One possible scenario is that these network jets are generated by magnetic reconnection. \citet{Axford1992} proposed that magnetic reconnection between the open field lines originating from network lanes and the adjacent closed loops may continuously inject materials upward to the fast solar wind. Network jets may be one form of such plasma ejections. Magnetic reconnection between emerging magnetic bipoles and overlaying unipolar fields often reveals inverted ``Y$"$-shape structures \citep[e.g.][]{Shibata2007}. However, only a few network jets show inverted ``Y$"$-shape structures \citep{Tian2014}. Another idea was proposed by \citet{Cranmer2015}, who suggested that these jets are driven by the upward forces associated with outward propagating Alfv\'{e}n waves. Some network jets are likely on-disk counterparts and heating signatures of the type-II chromospheric spicules, as demonstrated by \citet{Pereira2014} and \citet{Rouppe2015}. However, the apparent velocities of network jets are about twice as large as those of the type-II spicules. \citet{Tian2014} suggested that only some apparent motions correspond to mass flows, while others may result from effects such as thermal evolution, rapid ionization in a dynamic heating environment, or propagation of shocks. Through numerical simulations, \citet{DePontieu2017} and \citet{Chintzoglou2018} have clearly demonstrated that some fast apparent motions are indeed caused by the rapid propagation of heating fronts. It is worth mentioning that recently \cite{Yang2018} simultaneously produced a fast TR jet and a classical chromospheric spicule through a numerical simulation of magnetic reconnection between the open network fields and horizontally advected closed fields, which do not support the above-mentioned connection of network jets to spicules.

Based on a spectroscopic observation with the Solar Ultraviolet Measurements of Emitted Radiation (SUMER) instrument \citep{Wilhelm1995} on board the Solar and Heliospheric Observatory (SOHO) spacecraft, \cite{Teriaca2004} suggested a possible link between the occurrence of EEs and macro spicules. The latter are likely clusters of network jets. Due to the low spatial resolution and the lack of imaging capability, it is difficult to use SUMER observations to investigate the relationship between EEs and network jets.

In this paper, we first identify EEs from the Si~{\sc{iv}} 1393.755 {\AA} line profiles, and then study the temporal and spatial relationship between different types of EEs and network jets using simultaneous spectroscopic and imaging observations of IRIS. Our investigation suggests that some EEs are related to network jets while others are not.

\section{Observation} \label{sec:obser}

IRIS performed a very large sit-and-stare observation of a coronal hole boundary region from 19:56 UT to 23:26 UT on 2013 Oct 8. In this observation IRIS pointed at (249$^{\prime\prime}$, 283$^{\prime\prime}$), with a roll angle of 90 degrees. The spatial pixel size was $\sim$0.$^{\prime\prime}$167 per pixel for both the spectral images and slit-jaw images. The cadence of the spectral observation was $\sim$32 s. We use the Si~{\sc{iv}} 1393.755 {\AA} spectra to identify EEs. The spectral dispersion in the far ultraviolet was $\sim$0.013 {\AA} per pixel. The absolute wavelength calibration was performed by assuming that the chromospheric Fe~{\sc{ii}} 1392.817 {\AA} line has an average Doppler shift of zero. 
Slit-jaw images were taken with a cycle of 3 slots at a cadence of 95 s. The first 2 slots are 1400 {\AA} and 2796 {\AA} exposures. The third slot has one 2832 {\AA} exposure for every five 1330 {\AA} exposures, resulting in a regular cadence of 570 s for the 2832 {\AA} filter and an average cadence of 114 s for the 1330 {\AA} filter. The exposure time of both spectral observation and imaging observation were $\sim$30 s. Similar network jets are seen in both the SJI 1330 {\AA} and 1400 {\AA} passbands. However, they have stronger emission and thus are easier to be identified in the SJI 1330 {\AA} images. So here we use the SJI 1330 {\AA} images for the identification of network jets. Each SJI image has been enhanced using an unsharp masking technique: we first smooth the original image over 6$\times$6 pixels, then subtract the smoothed image from the original one, and finally add the residual to the original image.

\begin{figure*} 
\centering {\includegraphics[width=\textwidth]{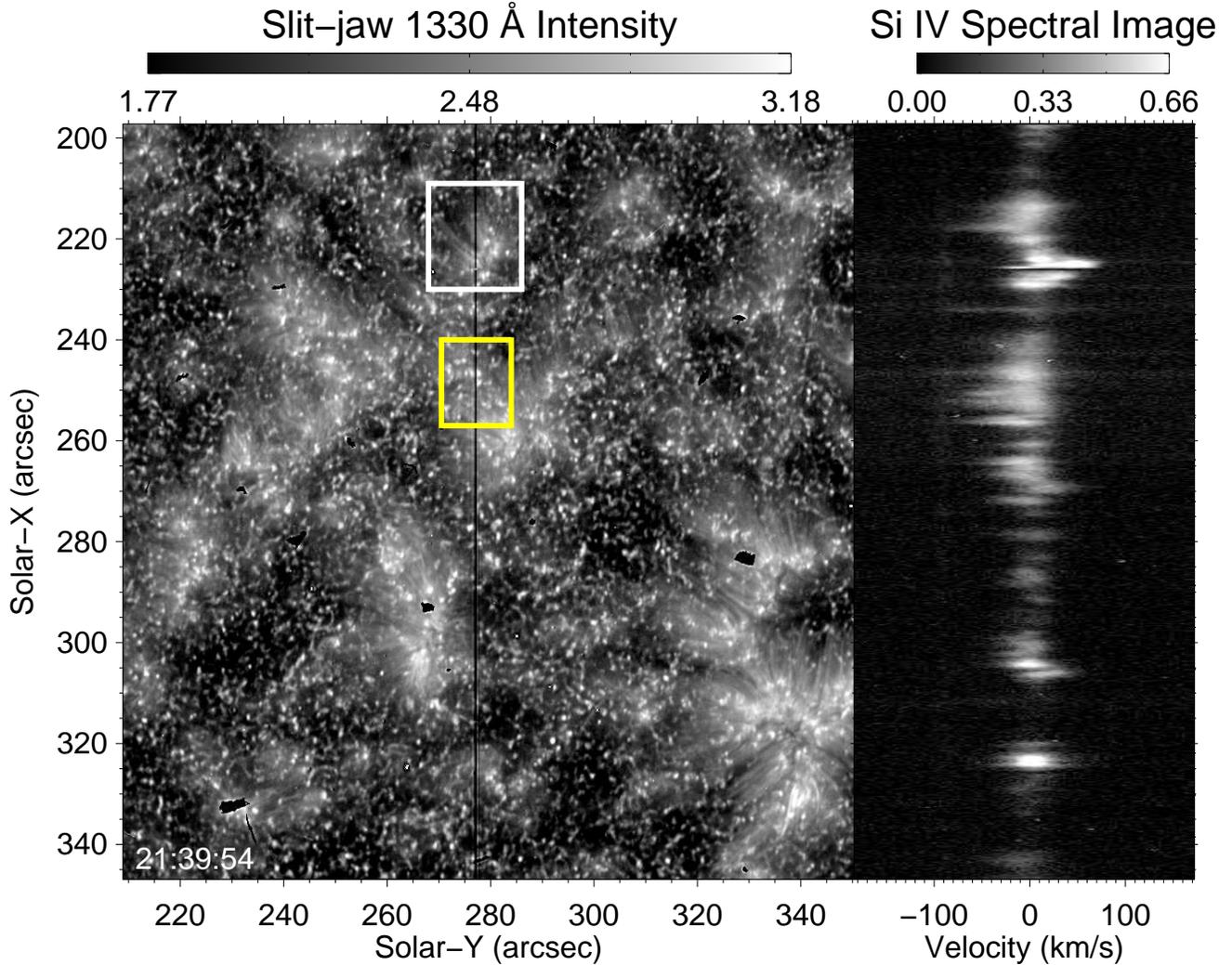}} 
\caption{Left: An unsharp masked SJI 1330 {\AA} image taken around 21:39:54 UT on 2013 Oct 8.  The white box outlines a region with network jet activity. The yellow box shows the field of view in Figure~\ref{f8}(a). The vertical black line in the middle marks the location of the slit. Right: Si~{\sc{iv}} 1393.755 {\AA} spectral image taken at the same time along the vertical slit. The intensities are shown in arbitrary unit. } \label{f1}
\end{figure*}

Figure ~\ref{f1} presents the SJI 1330 {\AA} image and Si~{\sc{iv}} 1393.755 {\AA} spectral image taken around 21:47:50 UT. The slit crosses several jet-like structures around the network lanes, especially within the region outlined by the white box. Note that due to the weak emission these network jets are better seen in movies than in still images. The Si~{\sc{iv}} spectra at the same location clearly reveal bi-direction flow structures, which are similar to those reported by \citet{Innes1997} and likely EEs. Thus, this dataset provides us a good opportunity to study the relationship between the TR network jets and EEs.

\section{Identification of Explosive Events} \label{sec:identification}

\begin{figure*} 
\centering {\includegraphics[width=\textwidth]{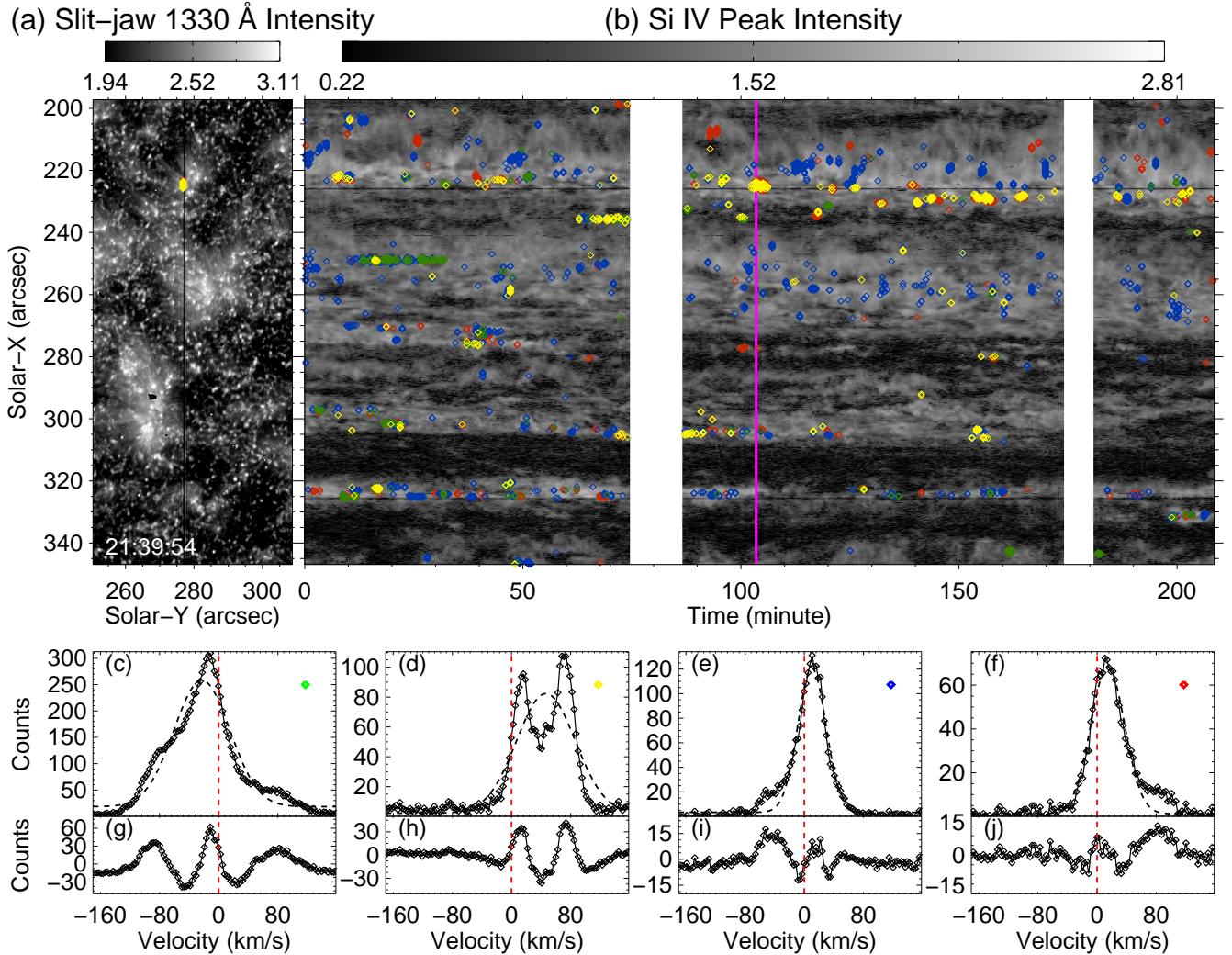}} 
\caption{ Identification of different types of EEs. (a): An unsharp masked SJI 1330 {\AA} image taken at 21:39:54 UT. (b): Temporal evolution of the Si~{\sc{iv}} line intensity. The start time is 19:56:58 UT. The vertical purple line corresponds to the time shown in panel (a). The white vertical bands indicate the periods when IRIS passed through the South Atlantic Anomaly (SAA), during which the data was significantly contaminated by high-energy particles. Different types of identified EEs are marked in panels (a) and (b). The green, blue, and red diamonds represent EEs with emission enhanced at both wings, enhanced at only the blue wing, enhanced at only the red wing, respectively. The yellow diamonds represent EEs with two emission peaks. (c)--(f): Examples of the four types of EEs. The colors of the isolated diamonds in panels (c)--(f) indicate the types of EEs. The black dashed lines represent the single Gaussian fits. (g)--(j): Residual profiles obtained by subtracting the single Gaussian fits from the original profiles in panels (c)--(f). The red dashed lines in panels (c)--(j) indicate the rest wavelength position of Si~{\sc{iv}} 1393.755 {\AA}. An online animation associated with panels (a)-(b) shows the 1330 {\AA} images, Si~{\sc{iv}} line intensities and identified EEs during 19:56:58 UT -- 23:24:26 UT. } \label{f2}
\end{figure*}

After visual inspection of many Non-Gaussian line profiles, we characterized four types of EEs based on the Si~{\sc{iv}} 1393.755 {\AA} line profiles. These profiles are either enhanced at both the blue and red wings (type I), show two peaks with comparable magnitude (type II), are enhanced only at the blue wing (type III) or red wing (type IV). We present the typical line profiles of different types of EEs in Figure ~\ref{f2}. Then we use a modified version of a detection method originally developed by \citet{Huang2017} to identify EEs. We briefly introduce our method here:

(1) We first filter out pixels polluted by cosmic rays, then perform a running average of the line profiles over three adjacent spatial pixels along the slit to increase the signal-to-noise ratio.

(2) Line profiles with peak intensities less than 30 counts or signal-to-background ratios (S/B) less than 10 are excluded as well. The S/B is defined as the peak intensity divided by the continuum intensity at the far wings \citep[see][]{Huang2017}.

(3) All the remaining line profiles are fitted with a single Gaussian function, and then the fitting result is subtracted from the original line profile to obtain the residual spectrum.

(4) We identify all local peaks of the residual spectrum. Each local peak is fitted with a single Gaussian function to obtain the intensity, width, and Doppler shift. If the intensity of one local peak is higher than 10$\%$ of the peak intensity of the original line profile or larger than 30 counts, width of the local peak is greater than 30$\%$ of the original line width, and the Doppler shift of the local peak is larger than 40 km~s$^{-1}$ at the blue or red wing, the local peak can be treated as an enhanced component at the blue or red wing. If a line profile has an enhanced component  at the blue or red wing, it will be identified as a candidate of EE.

(5) Then we examine the original line profiles of all identified EE candidates manually. If a line profile has enhanced components at both wings, with an obvious emission component in between, it will be identified as an EE with both wing enhancements. If a line profile exhibit two peaks with comparable magnitude, and the intensity of the weaker peak is higher than 50\% of the intensity of the stronger peak, the line profile will be identified as an EE with two peaks. If a line profile has an enhanced component only at the blue or red wing, while the intensity of the wing enhancement is lower than 50\% of the peak intensity of the line profile, it will be identified as an EE with only the blue or red wing enhancement.



\begin{table*}[]
\centering
\caption{Result of EE identification}
\label{t1}
\begin{tabular}{c|c}
\hline
\begin{tabular}[c]{@{}c@{}}Type of EEs\end{tabular}  & Number of events (Percentage)  \\ \hline
\begin{tabular}[c]{@{}c@{}}EEs with enhancement at both wings\end{tabular}  & 418 (13.7 \%)  \\ \hline
\begin{tabular}[c]{@{}c@{}}EEs with two comparable emission peaks\end{tabular}    & 465 (15.3 \%)   \\ \hline
\begin{tabular}[c]{@{}c@{}}EEs with only blue wing enhancement\end{tabular} & 1650 (54.1 \%) \\ \hline
\begin{tabular}[c]{@{}c@{}}EEs with only red wing enhancement\end{tabular}  & 514 (16.9 \%)  \\ \hline
\end{tabular}
\end{table*}

We have identified 3047 EEs in total, and the numbers and fractions of different types of EEs are shown in Table ~\ref{t1}. Figure ~\ref{f2} presents the result of identification. EEs are marked on the SJI 1330 {\AA} image and the Si~{\sc{iv}} intensity image. We find that almost all the EEs are located in the large-scale bright lane-like structures seen in the SJI 1330 {\AA} image. These relatively bright and steady network-like patterns in the Si~{\sc{iv}} intensity image mark the network lanes. This has been already found previously \citep[e.g.][]{Dere1994}. There are more EEs with only blue wing enhancement compared to EEs with only red wing enhancement, and a similar result was obtained by \citet{Teriaca2004}. Examples of different types of EEs identified in this dataset are also presented in the figure. Figure ~\ref{f2} shows that at least some EEs are associated with network jets. In the following, we investigate the relationship between the identified EEs and network jets.

\section{Results and Discussion} \label{sec:examples}

\begin{figure*} 
\centering {\includegraphics[width=\textwidth]{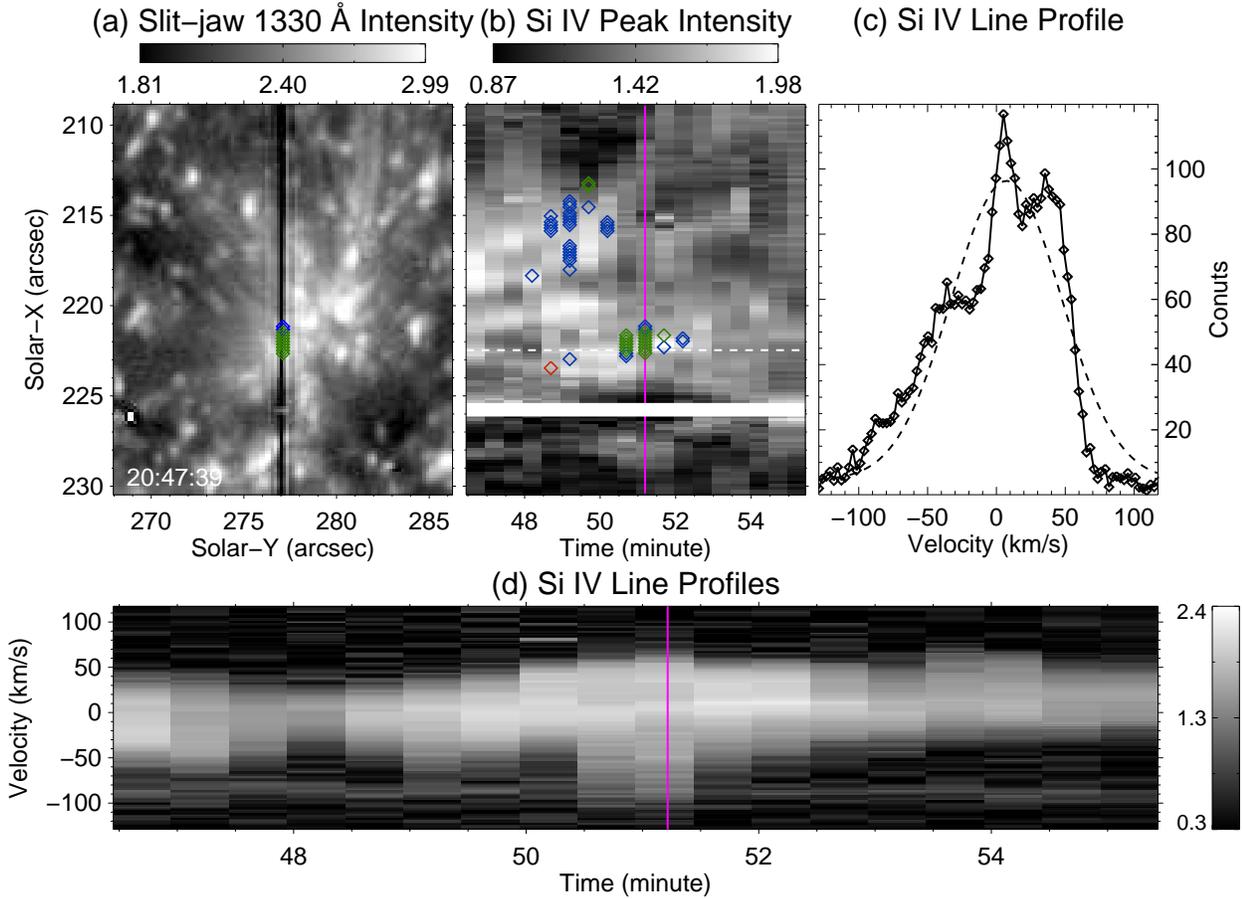}} 
\caption{ Example of EEs with both wing enhancements. (a): An unsharp masked SJI 1330 {\AA} image (logarithmic scale) taken around 20:47:39 UT. The FOV is the same as the white box outlined in Figure ~\ref{f1}. (b): Temporal evolution of the Si~{\sc{iv}} peak intensity. The vertical purple line indicates the time shown in (a). The white horizontal bar marks 3 pixels centered at the fiducial line. The green, blue, and red diamonds represent identified EEs with emission enhanced at both wings, enhanced at only the blue wing, and enhanced at only the red wing, respectively. The yellow diamonds represent EEs with two peaks. (c): Line profile at the intersection of the vertical purple line and the horizontal dashed line shown in (b). The dashed black line is the single Gaussian fit. (d): Wavelength$-$time diagram (logarithmic scale) of Si~{\sc{iv}} 1393.755 {\AA}. The vertical purple line indicates the time shown in (a).
} \label{f3}
\end{figure*}

\begin{figure*} 
\centering {\includegraphics[width=\textwidth]{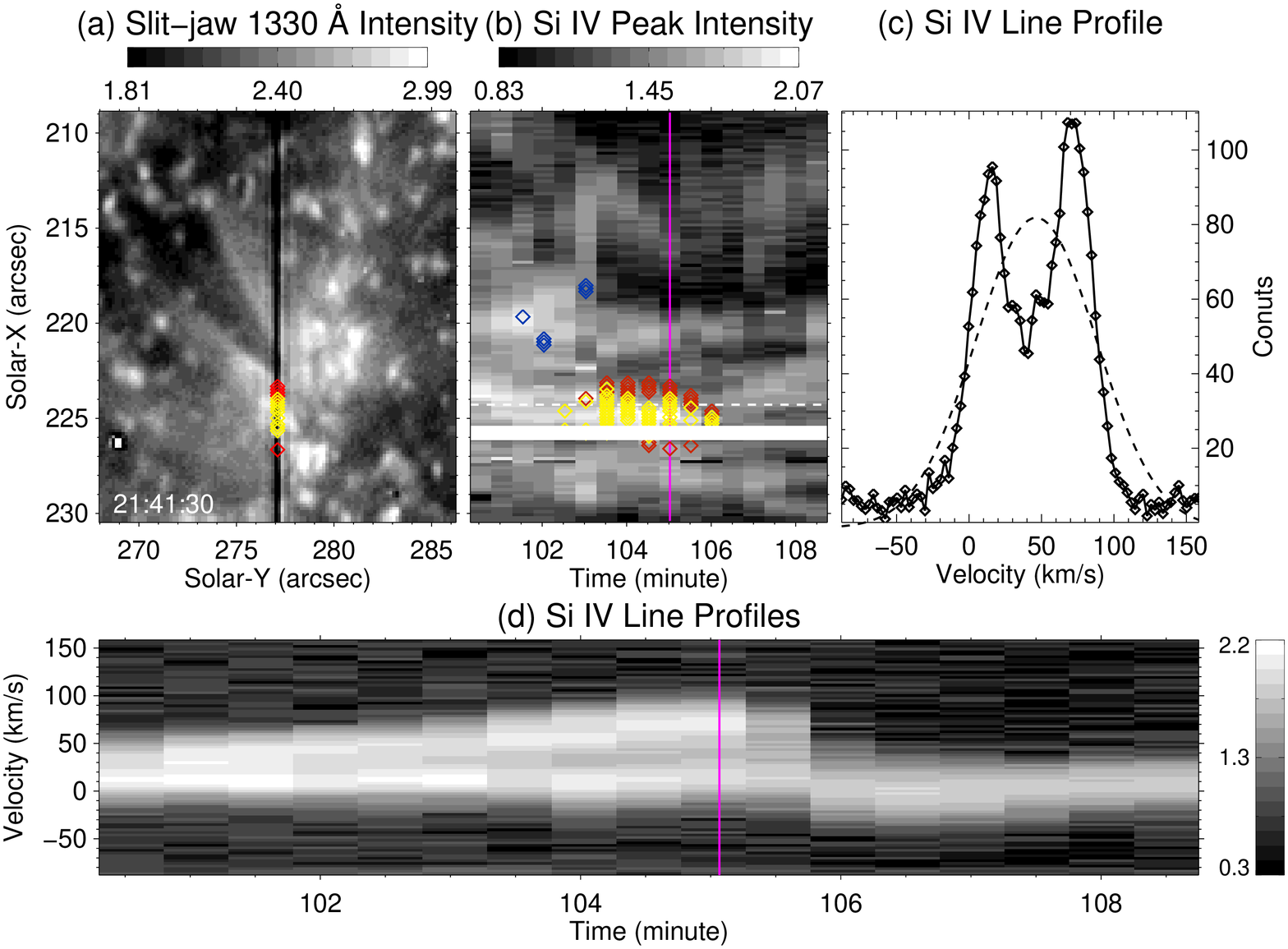}} 
\caption{ Example of EEs with two peaks. The types of images, curves, line styles and symbols are the same as those in Figure ~\ref{f3}. The line profile in (c) is the same as the one shown in Figure ~\ref{f2}(d).
} \label{f4}
\end{figure*}

Thanks to the simultaneous imaging and spectroscopic observation of IRIS, we can investigate the relationship between EEs identified from the spectral line profiles and the network jets observed from slit-jaw images.

\subsection{EEs at the Footpoints of Network Jets}

We notice that the slit crosses the footpoints of some network jets. We present the line profile of one EE at the footpoints of the jets in Figure~\ref{f3}. It clearly shows a non-Gaussian shape with an enhancement at both the blue and red wings. The wing enhancement is also obvious from the wavelength$-$time diagram. The enhancement at both wings might correspond to bi-directional reconnection outflows. Figure ~\ref{f4} presents examples of EEs with two peaks. They are located around the footpoints of some network jets as well. The line profile of one such EE is presented in the figure. From the wavelength$-$time diagram, we see only one peak at the beginning. A second peak appears around 21:37:16 UT ($\sim$100.7 minute), and afterwards the distance between the peaks becomes larger and larger. The second peak disappears around 21:42:10 UT ($\sim$105.6 minute), and only one peak around the rest wavelength of Si~{\sc{iv}} 1393.755 {\AA} remains. Interestingly, we find that the two peaks are both red-shifted. It seems unlikely that these two peaks correspond to the bidirectional reconnection jets in a simple geometry of magnetic reconnection. More likely, these peaks are related to the motions of plasmoids in a scenario of fast magnetic reconnection driven by plasmoid instability \citep[e.g.,][]{Innes2015,Li2019,Ni2015,Ni2016}. It is also possible that the two flows originate from a reconnection site that is accelerating downwards. We have to mention that in our dataset there are also EEs with one peak at the blue wing and the other one at the red wing. 

As mentioned above, EEs with enhanced emission at both wings (shown in Figure ~\ref{f3}) and those with two peaks (shown in Figure ~\ref{f4}) may both result from magnetic reconnection. Actually, \citet{Innes2015} reproduced both of these two types of EEs through a numerical simulation of magnetic reconnection driven by plasmoid instability. They found that the angle between the LOS and the current sheet could affect the shapes of the observed line profiles significantly. It is also possible that the emission enhancements at both wings are simply caused by bi-directional reconnection jets, and that the line core emission is related to the heated background plasma \citep{Ding2011, Heggland2009}. If the enhancements at both wings are much stronger than the background emission, the line profile could also show two peaks. In any case, the locations of these EEs likely mark the reconnection sites. Considering that some of these EEs are located at the footpoints of network jets, our observations support a scenario that these jets are driven by intermittent reconnection in the chromospheric network. A scenario of this continuous reconnection ultimately driving plasma upwards has been first suggested by  \citet[]{Axford1992} in the context of the acceleration of the fast solar wind. 

\begin{figure*} 
\centering {\includegraphics[width=\textwidth]{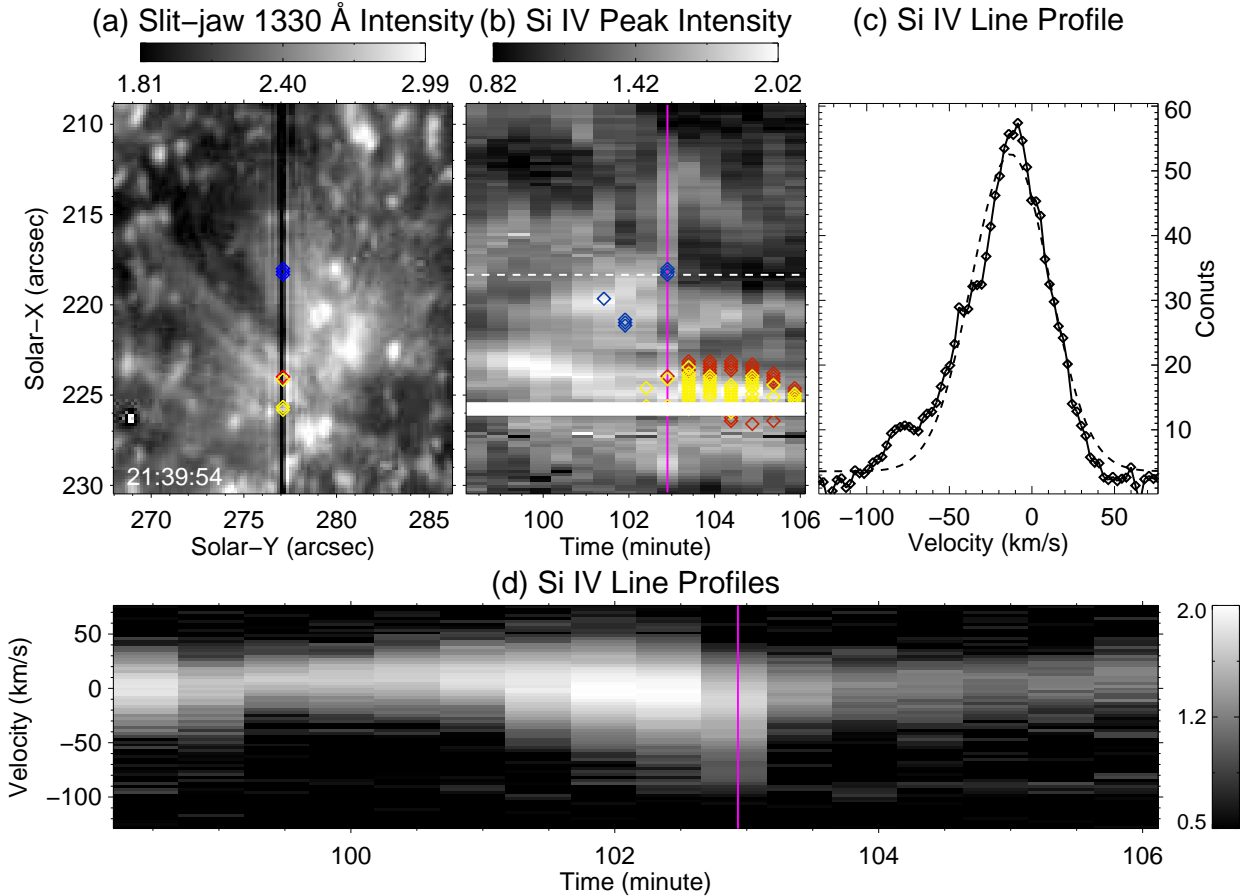}} 
\caption{Example of EEs with blue wing enhancement. The types of images, curves, line styles and symbols are the same as those in Figure ~\ref{f3}.} \label{f5}
\end{figure*}

\begin{figure*}
\centering {\includegraphics[width=\textwidth]{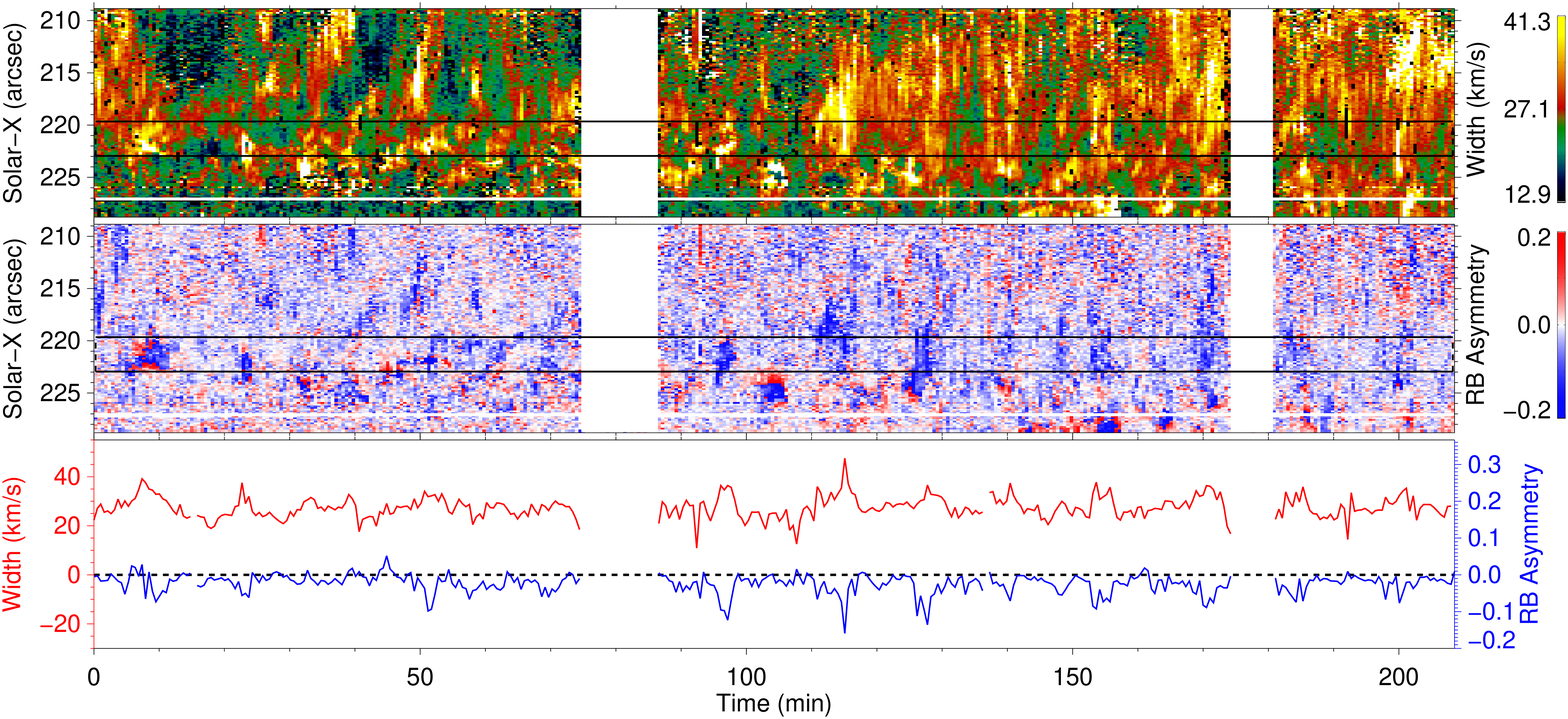}} 
\caption{Top: Temporal evolution of the Si~{\sc{iv}} line width obtained from the single Gaussian fit. Middle: Temporal evolution of the Si~{\sc{iv}} line profile asymmetry in the velocity range of 40--120 km~s$^{-1}$. The white vertical bands in the top and middle panels indicate the periods when IRIS passed through the SAA. Bottom: Temporal evolution of Si~{\sc{iv}} line width (red line) and profile asymmetry in the range of 40--120 km~s$^{-1}$ (blue line, in unit of percentage) averaged over the region between the two black lines in the top and middle panels. The dashed line indicates the absence of blue/red wing asymmetry, i.e., the blue and red wings in the range of 40--120 km~s$^{-1}$ are equally strong.
} \label{f6}
\end{figure*}

\subsection{EEs on Network Jets}

We can see that the slit crosses some faint network jets (around the blue diamonds) in Figure ~\ref{f5}. There are some compact small brightenings at the footpoints of the jets, which are located to the right of the slit. At the slit location we have identified some EEs with only the blue wing enhancement, and a typical line profile is shown in the figure. For the sake of simplicity we name them ``EEs on network jets''. The blueshifted component can also be identified from the wavelength$-$time diagram. The blueshifted component only lasts for $\sim$30 s, much shorter than the lifetime of the EE shown in Figure ~\ref{f4}. We have to mention that in reality there should be more line profiles with blue wing enhancement, as many jets appear to cross the slit around this time. However, the blue wing enhancements of many profiles are too weak and do not pass our identification criteria described in Section 3. Thus, these line profiles cannot be identified as EEs. 

Figure ~\ref{f6} presents the temporal evolution of the Si~{\sc{iv}} line width obtained from a single Gaussian fit and profile asymmetry in the velocity range of 40--120 km~s$^{-1}$. The slit section roughly corresponds to that shown in Figure ~\ref{f5}. We use the method introduced by \citet{Tian2011} to derive the profile asymmetry. The spectral position which corresponds to the peak intensity is determined as the line centroid. Then we subtract the red wing from the blue wing, and normalize it to the peak intensity. Temporal evolution of these parameters averaged along the slit over a 3$^{\prime\prime}$$-$segment is also presented. There is a clear positive correlation between the line width and blue wing enhancement, suggesting that a large fraction of the network jets are real upward mass flows \citep{Samanta2015}. Such a correlation has been previously reported for coronal upflows \citep{DePontieu2010, Tian2011, Tian2011b, Tian2012}. Our result suggests a similar scenario in the transition region.  
Thus, these EEs likely result from a superposition of the high-speed network jets on the TR background. Individual examples of the blue wing enhancement caused by network jets have been previously presented by \cite{Tian2014} and \citet{Rouppe2015}. 

We also find that the locations of EEs enhanced only at the blue wing and red wing are generally separated in Figure ~\ref{f2}, and this spatial separation has been previously found by \citet{Dere1989}. The online movie associated with Figure ~\ref{f2} shows that EEs enhanced only at the blue wing are mostly associated with the higher parts rather than the footpoints of network jets. And EEs enhanced only at the red wing are often observed around the jet footpoints. Additionally, we find that the number of detected EEs with only red wing enhancement is smaller than those with only blue wing enhancement (see Table ~\ref{t1}). These results may be understood in a scenario of magnetic reconnection between emerging magnetic bipoles and overlaying magnetic fields, which produces small-scale jets in the chromosphere and TR \citep[e.g.,][]{Shibata2007,Moore2011,Yurchyshyn2013,Tian2018}. In such a scenario the downflow regions are located around the footpoints of the jets, while the upflow regions can extend longer along the jets. Another possible explanation is that the downward reconnection outflows are easier to slow down due to the larger density in the lower atmosphere, thus downflows cannot propagate as far as upflows. The EEs with only red wing enhancement may result from the superposition of these downflows on the background emission. If the slit does not cross a reconnection site, the observed line profiles may only reveal a blue or red wing enhancement. A two-peak line profile may also be expected if the wing emission enhancement is comparable to a weak background.

\begin{figure*} 
\centering {\includegraphics[width=\textwidth]{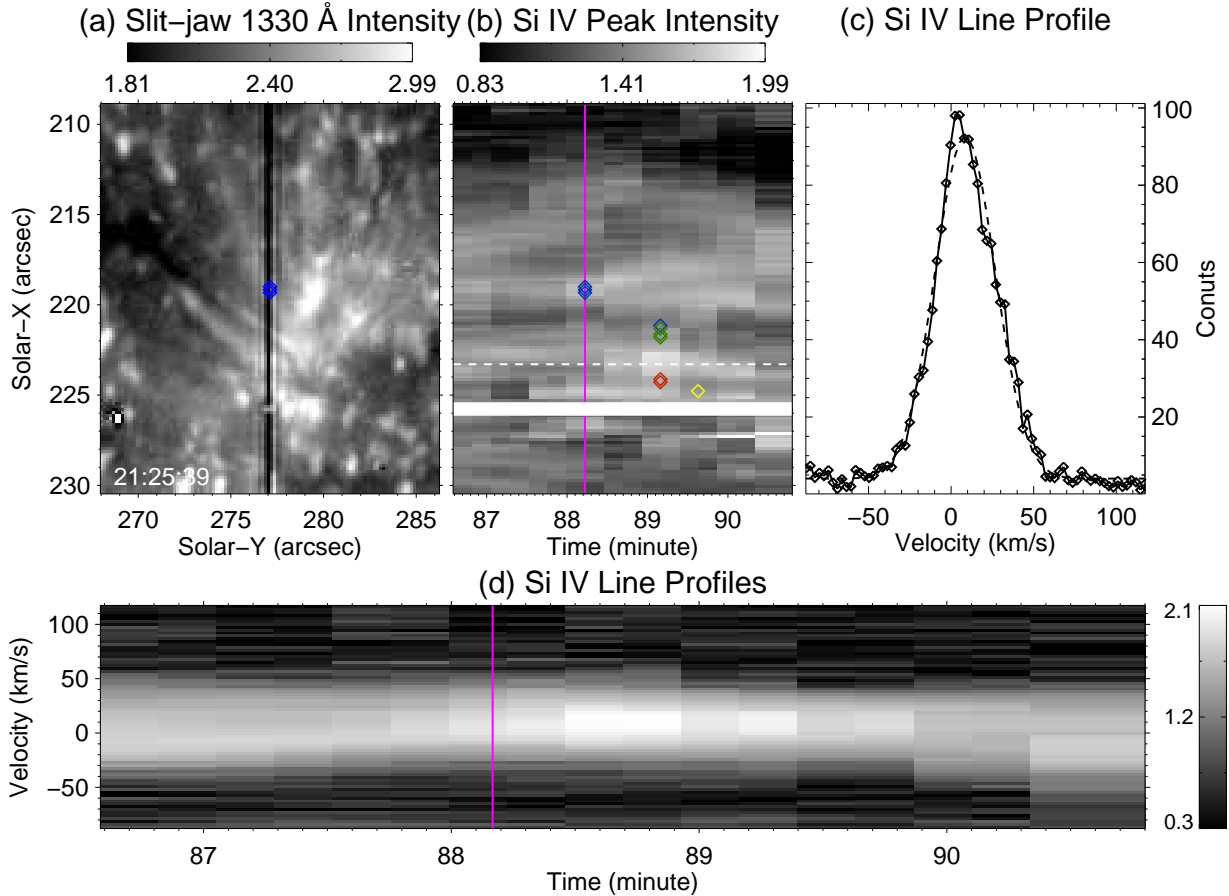}} 
\caption{Example of network jets showing no connection to EEs. The types of images, curves, line styles and symbols are the same as those in Figure ~\ref{f3}.} \label{f7}
\end{figure*}

Additionally, we find a few examples of network jets showing no obvious connection to EEs. One such example is presented in Figure ~\ref{f7}. We examine the line profile on the jet visible in the TR image, and find no clear deviation from a Gaussion distribution. The wavelength$-$time diagram also reveals no enhancement at the line wings. Possibly, these apparent motions are not caused by real mass flows, but by other effects such as the heating fronts found in the simulations of \cite{DePontieu2017} and \cite{Chintzoglou2018}. It is also possible that the enhancement at the line wings is too weak, or the Doppler velocity of the enhancement is lower than 40 km~s$^{-1}$. In such cases the line profiles may not be identified as EEs.

\begin{figure*} 
\centering {\includegraphics[width=\textwidth]{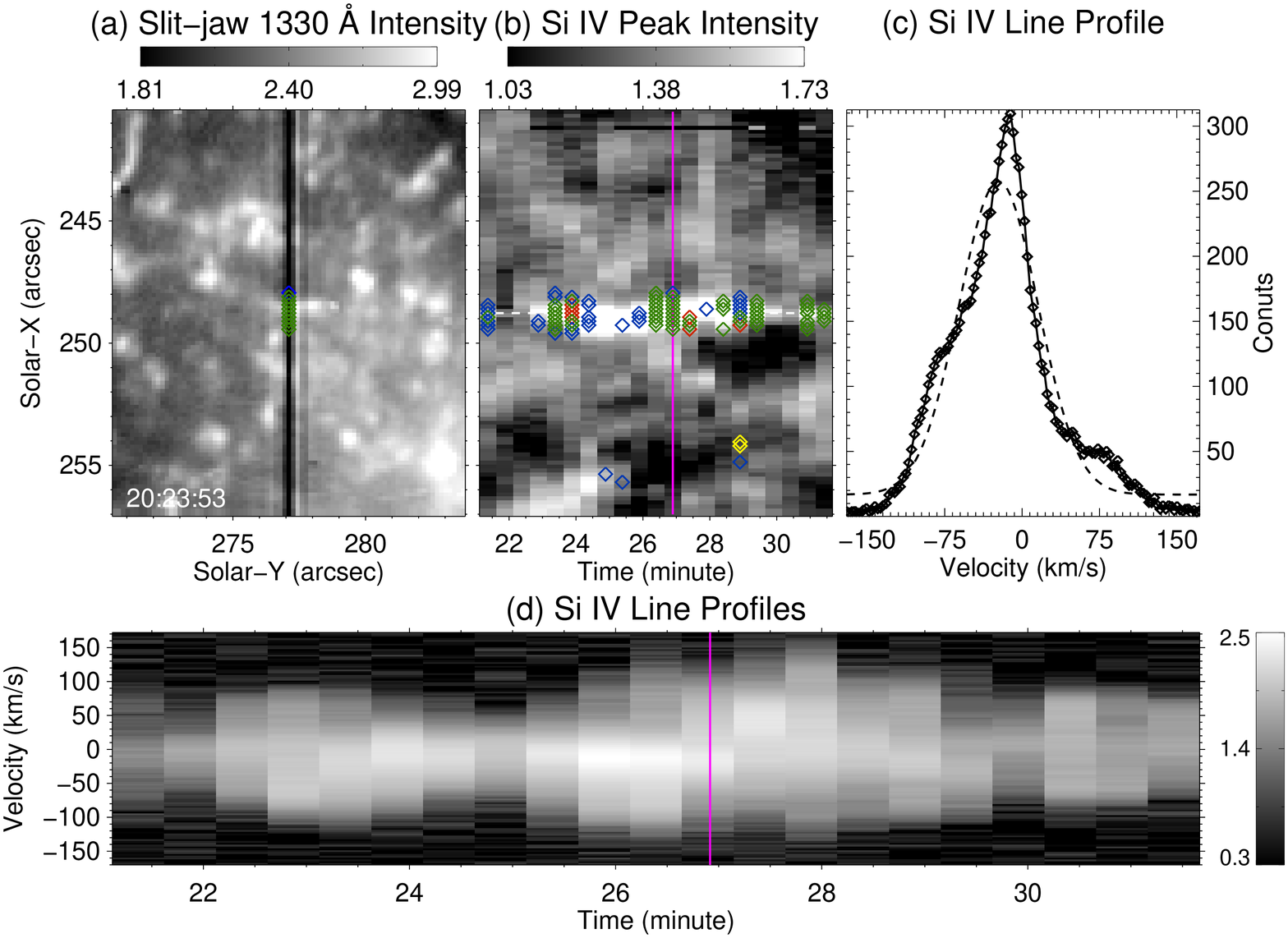}} 
\caption{ Example of EEs that are related to transient bright points in the SJI 1330 {\AA} images. The types of images, curves, line styles and symbols are the same as those in Figure ~\ref{f3}. The FOV in (a) is the same as the yellow box outlined in Figure ~\ref{f1}. The line profile in (c) is the same as the one shown in Figure ~\ref{f2}(c). An online animation associated with panels (a)-(c) shows the 1330 {\AA} images, Si~{\sc{iv}} line intensities and identified EEs during 20:19:08 UT -- 20:30:13 UT. } \label{f8}
\end{figure*}

\begin{figure*} 
\centering {\includegraphics[width=\textwidth]{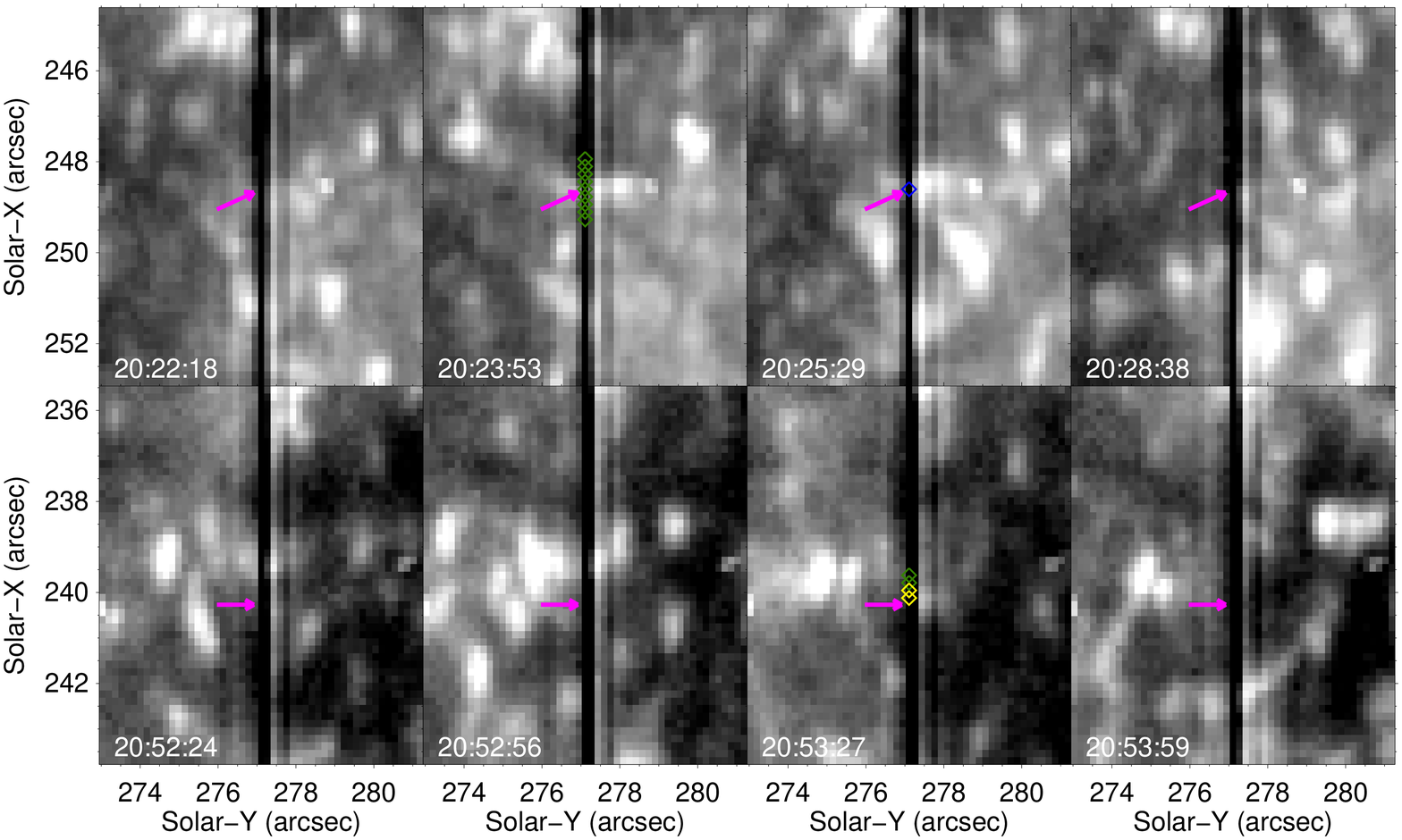}} 
\caption{Examples of EEs associated with transient bright points. The images have been enhanced through unsharp masking. The identified EEs are marked in each panel (green--enhanced at both wings; yellow--two peaks). Top: time evolution of the bright point related to the EEs shown in Figure ~\ref{f8}. Bottom: time evolution of a bright point related to EEs with both wing enhancements. The purple arrows indicate locations of the bright points on the slit.} \label{f9}
\end{figure*}

\subsection{EEs not Associated with Network Jets}

In addition, we find that some EEs with both wing enhancements or with two peaks are located around small-scale transient bright points in the slit-jaw images. One example of such EEs is presented in Figure ~\ref{f8}. Enhanced emissions at both the blue and red wings can also be identified from the wavelength$-$time diagram, and last for a few minutes. The enhancements at both wings are likely caused by bi-directional reconnection jets or motions of numerous plasmoids in a current sheet. There are no obvious jet-like structures in the SJI 1330 {\AA} images. Instead, these EEs are clearly related to transient small-scale bright points in the slit-jaw images. Figure ~\ref{f9} presents the time evolution of a bright point, which is clearly related to the EEs marked in Figure ~\ref{f8}. Another similar event is also presented in Figure ~\ref{f9}. 

These small-scale transient bright points might be small versions of UV bursts in the quiet sun regions. UV bursts are often characterized by stronger and wider Si~{\sc{iv}} line profiles, and are mostly not associated with visible extended jets in the TR images \citep[e.g.][]{Peter2014, Vissers2015, Tian2016, Young2018}. Our observed transient bright points may be the heating signatures of magnetic reconnection in the lower atmosphere, possibly fall into the category of quiet-Sun reconnection events recently described by \cite{Rouppe2016}, \cite{Nelson2017} and \cite{Shetye2018}.  The absence of jets may be caused by the projection effect, i.e., the reconnection jets propagate approximately along the LOS. It is also possible that the jets are too low in intensity or too small in size to be detected. 

\section{Summary}\label{sec:sum}

We have investigated the TR EEs and their relationship to network jets using simultaneous spectral and imaging observations from IRIS. Based on the properties of the identified EEs, we have categorized them into four groups: i) line profiles are enhanced at both wings, ii) line profiles show two peaks with comparable magnitude, iii)  line profiles show enhancement only at the blue, and iv) the red wing. 

Our analysis suggests that EEs in groups i) and ii) are mainly located at either the footpoints of network jets or transient compact brightenings that are not associated with network jets in SJI 1330 {\AA} images. The locations of these EEs likely mark the reconnection sites, and the spectral line profiles could be formed as a result of bidirectional reconnection outflows or motions of plasmoids in the reconnection current sheets. EEs in group iii) are mainly located on the network jets and away from the  jet foopoints. These EEs are obviously related to the upward propagation of the jets. EEs in group iv) are mainly located around the footpoints of network jets, likely related to the downward moving reconnection ouflows.  

These results suggest that some EEs are related to the birth or propagation of network jets, though there are still many network jets which reveal no signs of EEs. The presence of EEs at the footpoints of some network jets provides support to the reconnection driven scenario of network jets. Our analysis also suggests that some EEs are not connected to any visible network jets. Instead, these EEs are associated with small-scale transient bright points in the SJI 1330 {\AA} images, which  could be characterized as quiet-Sun UV bursts. 

\begin{acknowledgements}

This work is supported by NSFC grants 41574166, 11790304(11790300) and 11825301, and the Max Planck Partner Group program. Z.H. is supported by the Young Scholar Program of Shandong University, Weihai (2017WHWLJH07). IRIS is a NASA small explorer mission developed and operated by LMSAL with mission operations executed at NASA Ames Research center and major contributions to downlink communications funded by ESA and the Norwegian Space Centre. We thank Dr. Song Feng for helpful discussions.

\end{acknowledgements}

\end{document}